\documentclass[aps,prstab,reprint,twocolumn,groupedaddress,showpacs,amsmath,amssymb]{revtex4-1}
\usepackage{epsfig}
\usepackage{amsmath}
\usepackage{dcolumn}
\usepackage{bm}
\usepackage{esint}
\usepackage{xcolor}

\newcommand{\No}{$^{\mathrm{14}}$N$^{\mathrm{7+}}$}
\newcommand{\Ni}{$^{\mathrm{14}}$N$^{\mathrm{3+}}$}

\begin{document}


\title{Extension of Busch's Theorem to Particle Beams}


\author{L.~Groening and C.~Xiao}
\affiliation{GSI Helmholtzzentrum f\"ur Schwerionenforschung GmbH, Darmstadt D-64291, Germany}
\email[]{la.groening@gsi.de}
\author{M.~Chung}
\affiliation{Ulsan National Institute of Science and Technology, Ulsan 44919, Republic of Korea}
\email[]{mchung@unist.ac.kr}


\date{\today}

\begin{abstract}
In 1926, H.~Busch formulated a theorem for one single charged particle moving along a region with a longitudinal magnetic field~[H.~Busch, Berechnung der Bahn von Kathodenstrahlen in axial symmetrischen electromagnetischen Felde, Z. Phys. {\bf 81} (5) p. 974, (1926)]. The theorem relates particle angular momentum to the amount of field lines being enclosed by the particle cyclotron motion. This paper extends the theorem to many particles forming a beam without cylindrical symmetry. A quantity being preserved is derived, which represents the sum of difference of eigen-emittances, magnetic flux through the beam area, and beam rms-vorticity multiplied by the magnetic flux. Tracking simulations and analytical calculations using the generalized Courant--Snyder formalism confirm the validity of the extended theorem. The new theorem has been applied for fast modelling of experiments with electron and ion beams on transverse emittance re-partitioning conducted at FERMILAB and at GSI.
\end{abstract}


\maketitle


In 1926, H.~Busch applied the preservation of angular momentum for systems with cylindrical symmetry to a charged particle moving inside a region with magnetic field $\vec{B}$~\cite{Busch,Reiser,Tsimring}. Using conjugated momenta, the magnetic field strength is intrinsically included into the equations of motion. In linear systems, the normalized conjugated momenta $p_x$ and $p_y$ are related to the derivatives of the particle position coordinates $(x,y)$ w.r.t. the main longitudinal direction of motion $\vec{s}$ through
\begin{flalign}
\label{canmoms_x}
& p_x\,:=\,x'+\frac{\mathcal{A}_x}{(B\rho)}\,=\,x'-\frac{yB_s}{2(B\rho)}\,,\\
\label{canmoms_y}
& p_y\,:=\,y'+\frac{\mathcal{A}_y}{(B\rho)}\,=\,y'+\frac{xB_s}{2(B\rho)}\,,
\end{flalign}
where $\vec{\mathcal{A}}$ is the magnetic vector potential with $\vec{B}=\vec{\nabla}\times\vec{\mathcal{A}}$, $B_s$ is the longitudinal component of the magnetic field, and $(B\rho)$ is the particle rigidity, i.e., its momentum per charge~$p/(qe)$, with $p$ as total momentum, $q$ as charge number, and $e$ as elementary charge.

Busch's Theorem~\cite{Busch,Reiser,Tsimring} states that the canonical angular momentum $\tilde{l}=xp_y-yp_x$ is a constant of motion that is written in cylindrical coordinates as
\begin{equation}
\label{Busch}
m\gamma r^2\dot{\theta}\,+\,\frac{eq}{2\pi}\psi\,=\,const.\,,
\end{equation}
where $\gamma$ is the relativistic factor, $r$ is the radius of transverse cyclotron motion around the beam axis, $\dot{\theta}$ is the corresponding angular velocity, and $\psi$ is the magnetic flux enclosed by this motion. Busch's Theorem for axially symmetric systems is on an invariant of motion of a single particle.

A general formulation of Eq.~(\ref{Busch}) has been derived in~\cite{SCF}, which is regarded as the generalized Busch's Theorem
\begin{equation}
\label{Busch_gen}
\oint_{\mathcal{C}}\vec{v}\cdot d\vec{C}\,+\,\frac{eq}{m}\psi\,=\,const.\,,
\end{equation}
i.e., the path integral of the stream of possible particle velocities $\vec{v}$ along a closed contour $\mathcal{C}$ confining a fixed set of possible particle trajectories plus the magnetic flux through the area enclosed by $\mathcal{C}$ is an invariant of the motion. Busch's Theorem of Eq.~(\ref{Busch}) is the special case of this generalized form for $\mathcal{C}$ being a circle of radius $r$.
This paper expresses an invariant through a sum of meaningful beam properties by re-formulating the invariance of the two eigen-emittances introduced in 1992 by A.J.~Dragt~\cite{Dragt}. This invariance holds strictly for the paraxial approximation and for mono-energetic beams as pointed out in~\cite{Floettmann}.

The two eigen-emittances $\tilde{\varepsilon}_{1/2}$ are equal to the two projected transverse beam rms-emittances $\tilde{\varepsilon}_{x/y}$, if and only if there are no correlations between the two transverse degrees of freedom~(planes). Eigen-emittances can be obtained by solving the complex equation
\begin{equation}
det(J\tilde{C}-i\tilde{\varepsilon}_{1/2}I)\,=\,0\,,
\end{equation}
where $I$ is the identity matrix and
\begin{equation}
\label{2nd_mom_matrix}
\tilde{C}=
\begin{bmatrix}
\langle x^2 \rangle &  \langle xp_x\rangle &  \langle xy\rangle & \langle xp_y\rangle \\
\langle xp_x\rangle &  \langle p_x^2\rangle & \langle yp_x\rangle & \langle p_xp_y\rangle \\
\langle xy\rangle &  \langle yp_x\rangle &  \langle y^2\rangle & \langle yp_y\rangle \\
\langle xp_y\rangle &  \langle p_xp_y\rangle & \langle yp_y\rangle & \langle p_y^2\rangle
\end{bmatrix}\,,
\end{equation}
\begin{equation}
J=
\begin{bmatrix}
0 &  1 &  0 & 0 \\
-1 &  0 &  0 & 0 \\
0 &  0 &  0 & 1 \\
0 &  0 & -1 & 0
\end{bmatrix}.
\end{equation}
Second moments $\langle uv\rangle $ are defined through a normalized distribution function $f_b$ as
\begin{equation}
\langle uv\rangle\:=\,\int\int\int\int f_b(x,p_x,y,p_y)\cdot uv\cdot dx\,dp_x\,dy\,dp_y
\end{equation}
and projected rms-emittances by~\cite{pu_uprime}
\begin{equation}
\tilde{\varepsilon}_u^2 \,:=\,\langle u^2\rangle\langle p_u^2\rangle - \langle up_u\rangle ^2\,.
\end{equation}
For two degrees of freedom, the two eigen-emittances can be calculated from~\cite{Xiao_prstab2013}
\begin{equation}
\label{eigen12}
\tilde{\varepsilon}_{1/2}=\frac{1}{2} \sqrt{-tr[(\tilde{C}J)^2] \pm \sqrt{tr^2[(\tilde{C}J)^2]-16\,det(\tilde{C}) }}\,.
\end{equation}
As the two eigen-emittances are preserved for the symplectic transformation~\cite{Dragt}, the sum of their squares is preserved as well, i.e.,
\begin{equation}
\label{sum_eigensq}
\begin{split}
\tilde{\varepsilon}_1^2\,+\,\tilde{\varepsilon}_2^2\,&=\,-\frac{1}{2}tr[(\tilde{C}J)^2]\,\\
& =\,\tilde{\varepsilon}_x^2\,+\,\tilde{\varepsilon}_y^2\,+\,2\,(\langle xy\rangle\langle p_xp_y\rangle-\langle yp_x\rangle\langle xp_y\rangle)\,\\
& =\,const.
\end{split}
\end{equation}
Using the definitions of $p_x$ and $p_y$ in Eq.~(\ref{2nd_mom_matrix}) together with Eq.~(\ref{eigen12}) and finally expanding Eq.~(\ref{sum_eigensq}) leads to
\begin{equation}
\begin{split}
\label{const}
& (\varepsilon_1-\varepsilon_2)^2\,+\,\left[\frac{AB_s}{(B\rho)}\right] ^2\,+\\
& \,2\frac{B_s}{(B\rho)}\left[\langle y^2\rangle\langle xy'\rangle - \langle x^2\rangle\langle yx'\rangle + \langle xy\rangle (\langle xx'\rangle - \langle yy'\rangle)\right]\,\\
& =\, const.\,
\end{split}
\end{equation}
where  $A\,:=\,\sqrt{\langle x^2\rangle\langle y^2\rangle -\langle xy\rangle ^2}$ is the rms-area of the beam divided by $\pi$. Quantities written as $\tilde{Q}$ are calculated from conjugated coordinates $(x,p_x,y,p_y)$ and those written as $Q$ are calculated from laboratory coordinates $(x,x',y,y')$; hence, $Q$ is obtained from $\tilde{Q}$ by substituting $(p_x,p_y)\rightarrow (x',y')$ in the expression defining $\tilde{Q}$. In the following, only laboratory coordinates are used, as the extended theorem will be applied to experiments that used these coordinates.

Equation~(\ref{const}) shows that changing both transverse eigen-emittances can be achieved through longitudinal magnetic fields as was proposed first in~\cite{Brinkmann_rep}, where the beam is created inside a region of longitudinal field being emerged afterwards into a region without a field. Successful experimental demonstration of this concept was reported in~\cite{Brinkmann_prstab}. The method has been applied to create very flat electron beams with aspect ratios of up to~100~\cite{Piot_prstab2006}. It was also proposed for ions being emerged from the solenoid field of an electron-cyclotron-resonance source to create beams of very low horizontal emittances that will allow for high-resolution spectrometers~\cite{Bertrand}. By placing a charge state stripper, i.e., changing $(B\rho )$ inside a solenoid, transverse emittance was adjustably transferred from one plane into the other one~\cite{Groening_prstab2011,Xiao_prstab2013,Groening_prl2014,Groening_IPAC15}.

\begin{figure}[hbt]
\centering
\includegraphics*[width=80mm,clip=]{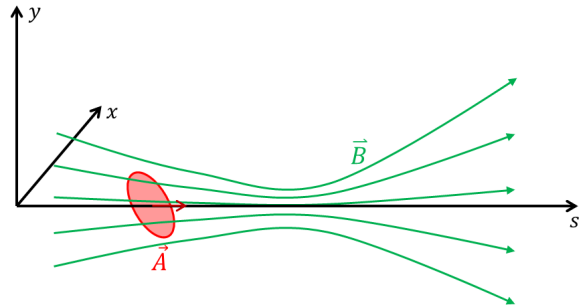}
\caption{Magnetic flux through the beam rms-area $\vec{A}$. Transverse field components do not contribute to the flux as they are perpendicular to the normal of the area.}
\label{flux}
\end{figure}

The first term of the left-hand side of Eq.~(\ref{const}) is the squared difference of the beam eigen-emittances. The second term is basically the square of the magnetic flux through the beam rms-area $\vec{A}$ as illustrated in Fig.~\ref{flux}.

In the following, it is shown that the essential part of the third term
\begin{equation}
\label{rms_Vorticity}
\mathcal{W}_A\,:=\,\langle y^2\rangle\langle xy'\rangle - \langle x^2\rangle\langle yx'\rangle + \langle xy\rangle (\langle xx'\rangle - \langle yy'\rangle)
\end{equation}
is the rms-averaged beam vorticity multiplied by the twofold beam rms-area.
We choose the ansatz assigning $\mathcal{W}_A$ to the rotation $(\vec{\nabla}\times )$ of the mean, i.e., averaged over $(x',y')$ space, beam angle $\vec{\bar{r'}}(x,y,s)$ being integrated over the beam rms-area, and finally multiplied by the twofold beam rms-area:
\begin{equation}
\mathcal{W}_A\,=2A \,\int\limits_A \,\left[\vec{\nabla}\times\vec{\bar{r'}}(x,y,s)\right]\, \cdot d\vec{A}
\end{equation}
being equivalent to
\begin{equation}
\label{ansatz}
\mathcal{W}_A\,=2A  \,\oint\limits_\mathcal{C} \vec{\bar{r'}}(x,y,s)\, \cdot d\vec{C}\,,
\end{equation}
where $\vec{\bar{r'}}(x,y,s) :=[\bar{x'}(x,y,s),\bar{y'}(x,y,s),1]$.
This ansatz is supported by the similarity of $\mathcal{W}_A$ to the first term of Eq.~(\ref{Busch_gen}). In continuum mechanics, the rotation of a media's velocity $(\vec{\nabla}\times\vec{v})$ is called the vorticity.

\begin{figure}[hbt]
\centering
\includegraphics*[width=80mm,clip=]{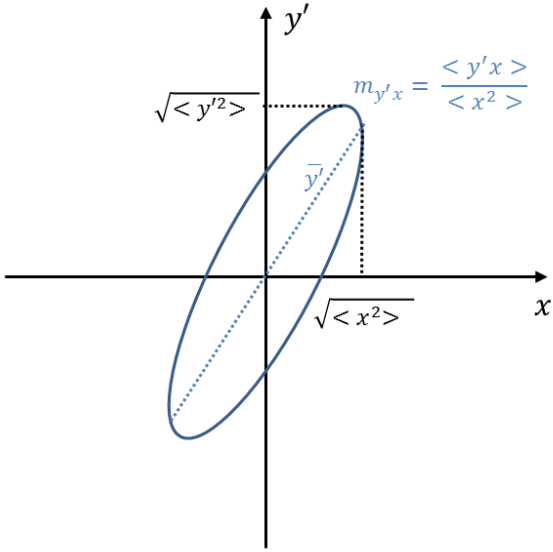}
\caption{Projection of the four-dimensional rms-ellipsoid onto the $(x,y')$~plane and the constant slope $(\partial \bar{y'}/\partial x)$.}
\label{ysx_plane}
\end{figure}

As $\mathcal{W}_A$ by construction is invariant under rotation by any angle in the $(x,y)$ plane, Eq.~(\ref{ansatz}) may be worked out for a beam with $\langle xy\rangle =0$ without loss of generality (imagine that prior to the calculation of $\mathcal{W}_A$ the beam is rotated around the beam axis by an angle that puts $\langle xy\rangle$ to zero).
For the following procedure, the resulting beam rms-area (divided by $\pi$) \mbox{$A=\sqrt{\langle x^2\rangle\langle y^2\rangle }$} is treated as being infinitesimally small in the paraxial approximation. Accordingly, the transverse components of $\vec{\bar{r'}}$ are expressed through the first terms of the Taylor series
\begin{align}
\bar{x'}(x,y)\,:=\,\bar{x'}(0,0)\,+\,\frac{\partial \bar{x'}}{\partial x}\cdot x\,+\,\frac{\partial \bar{x'}}{\partial y}\cdot y,\\
\bar{y'}(x,y)\,:=\,\bar{y'}(0,0)\,+\,\frac{\partial \bar{y'}}{\partial x}\cdot x\,+\,\frac{\partial \bar{y'}}{\partial y}\cdot y,
\end{align}
which turns into
\begin{align}
\label{mean_rs_a}
\bar{x'}(x,y)\,:=\,\frac{\langle x'x\rangle }{\langle x^2\rangle }x\,+\,\frac{\langle x'y\rangle }{\langle y^2\rangle }y,\\
\bar{y'}(x,y)\,:=\,\frac{\langle y'x\rangle }{\langle x^2\rangle }x\,+\,\frac{\langle y'y\rangle }{\langle y^2\rangle }y.
\label{mean_rs_b}
\end{align}
Figure~\ref{ysx_plane} illustrates as an example the constant slope $(\partial \bar{y'}/\partial x)$ of $\bar{y'}$ in the projection of the four-dimensional rms-ellipsoid onto the $(x,y')$~plane.
The path integral around the rms ellipse $x^2/\langle x^2\rangle + y^2/\langle y^2\rangle = 1$ can be done
by the following changes of variables:
$x = \sqrt{\langle x^2\rangle} \cos \theta$,  $y = \sqrt{\langle y^2\rangle} \sin \theta$,
and
\begin{equation}
d \vec{C} = \left( \frac{dx}{d\theta},  \frac{dy}{d\theta} \right) d\theta = \left( - \sqrt{\langle x^2\rangle} \sin\theta, \sqrt{\langle y^2\rangle} \cos\theta \right) d \theta\,.
\end{equation}
Therefore,
\begin{eqnarray}
&~& 2A  \,\oint\limits_\mathcal{C} \vec{\bar{r'}}(x,y,s)\, \cdot d\vec{C}\, \nonumber \\
&=&  2A \int_0^{2\pi} \left( \,\frac{\langle x'x\rangle }{\langle x^2\rangle }x\,+\,\frac{\langle x'y\rangle }{\langle y^2\rangle }y \right) \left( -  \sqrt{\langle x^2\rangle} \sin\theta \right) d \theta  \nonumber  \\
&+&  2A \int_0^{2\pi} \left( \,\frac{\langle y'x\rangle }{\langle x^2\rangle }x\,+\,\frac{\langle y'y\rangle }{\langle y^2\rangle }y\ \right) \left(  \sqrt{\langle y^2\rangle} \cos \theta \right)  d \theta \nonumber  \\
&=& \langle y'x\rangle \langle y^2\rangle  - \langle x'y \rangle  \langle y^2\rangle\,=\,\mathcal{W}_A\,,
\end{eqnarray}
which proves that the ansatz is correct.

For the time being, acceleration has not been included into the treatment. This can be done simply by multiplying Eqs.~(\ref{canmoms_x}) and (\ref{canmoms_y}) initially by \mbox{$p=m\gamma\beta c$}. $\beta$ is the longitudinal particle velocity normalized to the velocity of light~$c$. The extension of Busch's Theorem to beams including acceleration is
\begin{equation}
\label{const_accel}
(\varepsilon_{n1}-\varepsilon_{n2})^2\,+\,\left[\frac{eq\psi }{mc\pi}\right] ^2\,+\,\frac{4eq\psi\beta\gamma}{mc\pi}\,\oint\limits_\mathcal{C} \vec{\bar{r'}}\cdot d\vec{C}\,
=\, const.\,,
\end{equation}
where $\psi$ is the magnetic flux through the beam rms-area~$A$.
Analogue to the normalized emittance $\varepsilon_n:=\beta\gamma\varepsilon $, the normalized beam rms-vorticity is introduced as
\begin{equation}
\mathcal{W}_{An}\,:=\,\beta\gamma\mathcal{W}_A\,.
\end{equation}

\begin{figure}[hbt]
\centering
\includegraphics*[width=86mm,clip=]{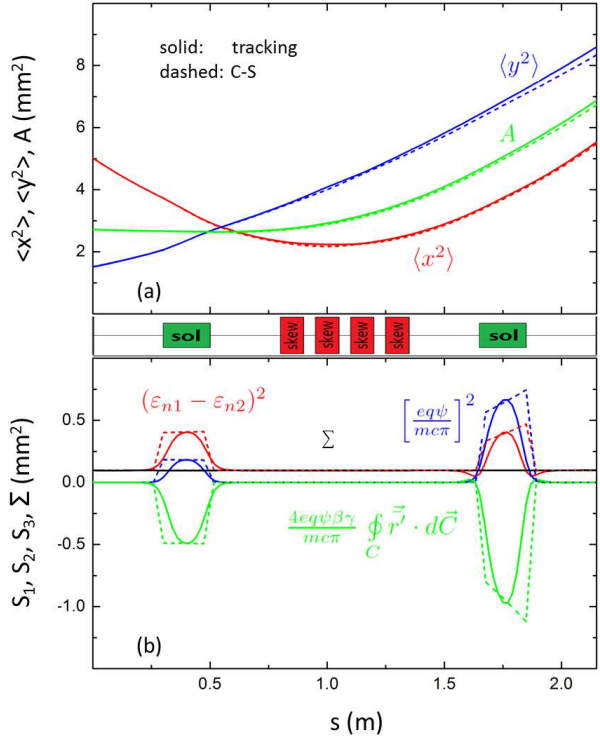}
\caption{(a): horizontal/vertical beam size (red/blue) and beam rms-area (green) along the beam line. (b): Summands of Eq.~(\ref{const_accel}) and their sum along the beam line. Results from tracking (C-S formalism) are plotted in solid~(dashed).}
\label{tracking}
\end{figure}

Tracking simulations using the BEAMPATH~\cite{BEAMPATH} code have been performed in order to verify Eq.~(\ref{const_accel}). The probe beam line (Fig.~\ref{tracking}) comprises a solenoid with an extended fringe field, a skewed quadrupole magnet quartet, and another extended solenoid. Figure~\ref{tracking} plots the beam widths, rms-area, the three summands of Eq.~(\ref{const_accel}), and their sum along the beam line. Additionally, the results from the application of the generalized Courant--Snyder (C--S) formalism for coupled lattices~\cite{Chung_prl2016} are plotted. In the latter, hard-edge solenoids with infinite short fringe field lengths have been assumed.
The three summands change exclusively along regions with a longitudinal magnetic field. Behind these regions, each of them gets back to the value it had prior to entering this region, respectively. The sum of the three beam properties remains constant in accordance to Eq.~(\ref{const_accel}).

At FERMILAB's NICADD photoinjector, flat electron beams were formed by first producing the beams at the surface of a photo cathode placed inside an rf-gun to which longitudinal magnetic field $B_s=B_0$ was imposed~\cite{Piot_prstab2006}. Along the subsequent region with \mbox{$B_s=0$}, the beam was accelerated to 16~MeV. Finally, correlations initially imposed by the magnetic exit fringe field of the rf-gun were removed by three skew quadrupole magnets. Equation~(\ref{const_accel}) equalizes the situation at the cathode surface at the left-hand side to the situation of the finally flat beam on the right-hand side~$(q=1)$
\begin{equation}
\label{piot1}
0\,+\,\left[\frac{eB_0A_0}{mc}\right]^2\,+\,0\,=\,(\varepsilon_{nf1}-\varepsilon_{nf2})^2\,+\,0\,+\,0\,,
\end{equation}
where $A_0$ is the beam rms-area at the cathode surface. The authors of~\cite{Piot_prstab2006} used the definitions~\cite{Kim}
\begin{align}
(\varepsilon^u_n)^2\,&:=\,\varepsilon_{nf1}\cdot\varepsilon_{nf2}\\
 \mathcal{L}\,&:=\,(eB_0A_0)/(2m\gamma\beta c)
\end{align}
resulting in
\begin{equation}
\varepsilon_{nf1}\,=\,\mathcal{L}\beta\gamma\pm\sqrt{(\mathcal{L}\beta\gamma)^2+(\varepsilon^u_n)^2}\,,
\end{equation}
of which only the upper sign gives a meaningful positive result. Re-plugging this expression for $\varepsilon_{nf1}$ into Eq.~(\ref{piot1}) leads to
\begin{equation}
\varepsilon_{nf1/2}\,=\,\pm\mathcal{L}\beta\gamma\,+\,\sqrt{(\mathcal{L}\beta\gamma)^2+(\varepsilon^u_n)^2}\,,
\end{equation}
being identical to their original expression~(Eq.~(1) of~\cite{Piot_prstab2006}).

At GSI, the EMittance Transfer EXperiment (EMTEX) transferred emittance from one transverse plane into the other one by passing the beam through a short solenoid~\cite{Xiao_prstab2013,Groening_prstab2011,Groening_prl2014,Groening_IPAC15}.
In the solenoid center, the ions charge state, i.e., their rigidity was changed by placing a thin carbon foil therein from \Ni ~to \No. Charge state stripping is a standard procedure used at several laboratories that deliver heavy or intermediate mass ions~\cite{Okuno_prstab,Scharrer_prab}. In front of the solenoid, the beam had no inter-plane correlations, and thus, the difference of rms-emittances was equal to the difference of eigen-emittances (mod.~sign). Since the solenoid was short, the beam area at the foil $A:=A_f$ can be approximated as constant during the beam transit through the solenoid.
Equation~(\ref{const}) relates the beam parameters in front of the solenoid ($B_s=0$, no correlations $\rightarrow\mathcal{W}_A =0,\,\,\varepsilon_{10}=\varepsilon_{x,3+},\,\,\varepsilon_{20}=\varepsilon_{y,3+}$) to those in front of the foil in the center of the short solenoid:
\begin{equation}
\begin{split}
& (\varepsilon _{10}-\varepsilon _{20})^2\,+\,0\,+\,0\,\\
& =\,(\varepsilon _{1f}-\varepsilon _{2f})^2\,+\,\left[\frac{A_f B_s}{(B\rho )_{3+}}\right]^2\,+\,\frac{2B_s}{(B\rho )_{3+}}\mathcal{W}_{Af}\,,
\end{split}
\end{equation}
where the index $f$ refers to the location of the foil.
The entrance fringe field of the solenoid causes the rms-vorticity
\begin{equation}
\label{def_kappa_3+}
\mathcal{W}_{A_f} = \Delta\mathcal{W}_A\,=\,-2\frac{B_s}{2(B\rho)_{3+}}A_f^2
\end{equation}
leading to
\begin{equation}
(\varepsilon _{x,3+}-\varepsilon _{y,3+})^2\,+\,0\,+\,0\,=\,(\varepsilon _{1f}-\varepsilon _{2f})^2\,-\,\left[\frac{A_f B_s}{(B\rho )_{3+}}\right]^2\,.
\end{equation}
Using the initial beam parameters of the experiment~\cite{Groening_prl2014}, $A_f=\sqrt{\varepsilon_x\beta_x\varepsilon_y\beta_y}=4.166\,\mathrm{mm}^2$, and the identity 1~mm~mrad~=~1~$\mu$m gives
\begin{equation}
\begin{split}
(\varepsilon _{1f}-\varepsilon _{2f})^2\,& =\,(\varepsilon _{x,3+}-\varepsilon _{y,3+})^2\,
+\,2.709~\mu \mathrm{m}^2\, \\
& =\,2.755~\mu \mathrm{m}^2\,.
\end{split}
\end{equation}
Equation~(\ref{const}) is re-used to relate the beam parameters that are just behind the foil but still at the center of the solenoid to those at the exit of the beam line, where \mbox{$B_s=0$} and the beam correlations have been removed again. Angular scattering in the foil is neglected. As the beam changed rigidity in the foil, $(B\rho )_{3+}$ must be properly replaced by $(B\rho )_{7+}$. However, second beam moments are not changed by the foil, i.e., $\mathcal{W}_A = \mathcal{W}_{A_f}$, right in front and right behind the foil. Accordingly,
\begin{equation}
\begin{split}
& (\varepsilon _{1f}-\varepsilon _{2f})^2\,+\,\left[\frac{A_f B_s}{(B\rho )_{7+}}\right]^2\,+\,\frac{2B_s}{(B\rho )_{7+}}\mathcal{W}_{Af}\,\\
& =\,(\varepsilon_{x,7+}-\varepsilon_{y,7+})^2\,+\,0\,+\,0\,,
\end{split}
\end{equation}
which by using Eq.~(\ref{def_kappa_3+}) and plugging in the values delivers
\begin{equation}
|\varepsilon_{x,7+}-\varepsilon_{y,7+}|\,=\,2.2~\mathrm{mm~mrad}
\end{equation}
fitting well the measured value of 2.0~mm~mrad (see Fig.~2 of~\cite{Groening_prl2014}).

The many particle pendant to Busch's Theorem on a single particle has been derived without requiring cylindrical symmetry but with including acceleration of the beam. It introduces the property of beam rms-vorticity and relates the beam's difference of eigen-emittances (i.e., intrinsic anisotropy), the magnetic flux through its area, and its rms-vorticity multiplied by the magnetic flux. Under the transport through coupled linear elements, the sum of these properties is preserved. The extended theorem was verified through tracking simulations and through application of the generalized C--S formalism for coupled dynamics. It was successfully used for quick and precise modelling of emittance re-partitioning experiments conducted at FERMILAB and at GSI, hence it is a powerful tool easily applicable to both electron and heavy ion beam lines or accelerators. The extended theorem significantly facilitates modelling and designing of devices for advanced emittance manipulations.

This research was partly supported by the National Research Foundation of Korea
(Grants No. NRF-2015R1D1A1A01061074 and No. NRF-2017M1A7A1A02016413).


\begin{thebibliography}{20}
\bibitem{pu_uprime}
Emittance definitions assume mono-energetic beams and refer to fixed position $s_0$ rather to fixed time $t_0$. The particle angle $u'$ and its transverse mechanical momentum \mbox{$P_u:=m\gamma v_u$} are related through $P_u=p\cdot u'$, where $p$ is the longitudinal mechanical momentum, which is the same for each particle.
\end{thebibliography}
\end{document}